# Infrared Radiation from Rough Surface

E.A. Ayryan [1], A.H. Gevorgyan [2], K.B. Oganesyan [1,3*]

[1] Joint Institute for Nuclear Research, LIT, Dubna, Moscow Region, Russia
[2] Far Eastern Federal University, Russky Island, Vladivostok, Russia
[3] Yerevan Physics Institute, Alikhanyan National Science Lab, Yerevan, Armenia

*bsk@yerphi.am

We consider generation of diffusive radiation by a charged particle passing through a random stack of plates in the infrared region. Diffusive radiation originates due to multiple scattering of pseudophotons on the plates. To enhance the radiation intensity one needs to make the scattering more effective. For this goal we suggest to use materials with negative dielectric constant .

## 1. Introduction

There is substantial interest in the development of infrared sources for applications to biophysics, nanostructures, material science and etc. Most lasers, for example Nd:YAG lasers, many fiber lasers and the most powerful laser diodes, emit near-infrared light. There are comparatively few laser sources for the mid- and far-infrared spectral regions. $CO_2$ lasers can emit at 10.6 μm and some other wavelengths in that region. Typical problems with laser crystals for solid-state mid-IR lasers are the limited transparency range of the host crystal and the tendency for fast multi-phonon transitions bypassing the laser transition; crystal materials with very low phonon energies are required. In recent years a lot of works were published where tunable infrared solid states and fiber lasers have operated at room temperature. Cryogenic lead-salt lasers were in earlier years often used for mid-infrared spectroscopy, but are now rivaled by quantum cascade lasers, which partly even achieve continuous-wave operation at room temperature.

Creation of compact inexpensive sources of radiation operating effciently in visible, UV, or soft X-ray domains is one of the most important directions in the development and investigation of Free-Electron Lasers. A short-wavelength radiation can be generated by a FEL using either a high-energy (multi-GeV) electron beam or undulators with a short period. One way of constructing short-period undulator-like media can be related with using Media with Periodically Modulated Refractive Index (MPMRI) [1-3]. MPMRI can be considered as a kind of a volume diffraction grating. The following two types of MPMRI can be used: (1) a gas-plasma medium with periodically varied density or degree of ionization [1] and (2) a spatially periodical solid-state superlattice-like (SLL) structure, which can be composed, e.g., of a series of layers of different materials with different refractive indices (see Ref. [2] and references therein). Note that a closely related but simpler effect of a stimulated transition emission was observed experimentally [4] in a scheme similar to that suggested in [1-2] but without media modulation. Free electron lasers and strophotrons [2] can be used as broadly tunable and very powerful sources of infrared light.

Coherent radiation from bunched electrons and prebunched FEL in far-infrared and the millimeter wavelength regions were reported in [5], ultrabroadband terahertz source and beamline based on coherent transition radiation are investigated in [6)], characterization and mitigation of coherent-optical-transition-radiation signals from a compressed electron beam are reported in [7]. The spontaneous radiation in multilayer systems is investigated in soft X-ray region [8], in EUV region [9], in soft X-ray and EUV regions [10] and in soft X-ray region [11].

The lack of electron beam driven devices is that usually they require large facilities [12-17].

In the present paper we discuss a possibility to create an infrared source using diffusive radiation [18,19]. The main underlying physical idea is following. One can make the average dielectric constant of a random stack of plates made of a material with negative dielectric constant and vacuum spacings between plates quite small. The pseudophoton momentum $k = \omega\sqrt{\varepsilon}/c$, where $\varepsilon$ is the average dielectric constant of the system, will correspondingly be small too. In such a system pseudophotons will scatter on inhomogeneties more effectively. Diffusive radiation is caused by multiple scattering of pseudophotons therefore its intensity will enhance in such a system. Note that the systems with near zero dielectric constant have many other interesting properties [20-22].

## 2. Radiation Intensity

A charged particle passing through a stack of plates placed in a homogeneous medium is known to be radiating electromagnetic waves. Radiation originates because of the scattering of electromagnetic field on the plates. Considering this problem theoretically earlier it was shown [18,19] that the spectral angular radiation intensity can be represented as a sum of two contributions

$$I = I_0 + I_D \tag{1}$$

where

$$I_0(\theta,\omega) = \frac{e^2}{2c} \frac{B(|k_0 - k\cos\theta|)\sin^2\theta}{\left(\gamma^{-2} + \sin^2\theta \cdot k^2/k_0^2\right)} \frac{\omega^2}{k_0^4 c^2} \tag{2}$$

and diffusive contribution is determined as

$$I_D(\theta,\omega) = \frac{5e^2\gamma^2}{2\varepsilon c}\frac{l_{in}^2(\omega)}{l^2(\omega)}\sin^2\theta \exp\left[-\left(\frac{l}{l_{in}}\right)^{1/2}\frac{1}{|\cos\theta|}\right] \quad (3)$$

Here $\theta$ is the observation angle, $k_0 = \omega/v$, v is the particle velocity, $k = \omega\sqrt{\varepsilon}/c$, $B$ is the correlation function of random dielectric constant field created by randomly located plates. Assuming that parallel plates with equal probability occupy any point of $z$ axis one finds correlation function as follows

$$B(q_z) = \frac{4(b-\varepsilon)^2 n\cdot\sin^2(q_z a/2)}{q_z^2}\frac{\omega^4}{c^4}. \quad (4)$$

where $n = N/L_z$ is the concentration of plates in the system, $a$ is their thickness, $b$ is their dielectric constant and $\varepsilon$ is the average dielectric constant of the system. In Eq.(3), $l$ and $l_{in}$ are average elastic and inelastic mean free paths of photon in the medium. Inelastic mean free path is mainly associated with the absorption of electromagnetic field in the medium. Elastic mean free path is associated with photon refraction on plates. It depends on the photon incidence angle on plates. In case of normally incident photons elastic mean free path is determined as follows

$$l = \frac{4k^2}{B(0) + B(2k)}. \quad (5)$$

Note that just this quantity enters into spectral angular intensity Eq.(3). Eqs.(3,5) are correct in the weak scattering limit $\lambda/l \ll 1$ and for observation angles $\theta = \pi/2 - \delta$, $\delta \gg (1/kl)^{1/3}$. Last restriction over angles appears because when $\theta = \pi/2$ pseudophotons move parallel to plates and $l = 0$ therefore the condition of weak scattering is failed. When the conditions of multiple scattering of electromagnetic field are fulfilled the diffusive contribution to the radiation intensity Eq.(3) is the main one because $I_D/I_0 \sim l_{in}/l$. As it is seen from Eq.(3) radiation intensity is determined by elastic and inelastic mean free paths of photon in the medium. In the next section we will investigate photon mean free paths in the infrared region in detail. It follows from Eq.(4) that when $ka \gg 1$, $B(2k)/B(0) \sim 1/(ka)^2 \ll 1$. Therefore the photon mean free path is [19]

$$l \approx \begin{cases} 4k^2/B(0), & ka \gg 1 \\ 2k^2/B(0), & ka \ll 1 \end{cases}.$$

In both cases $ka \gg 1$ and $ka \ll 1$ photon mean free path has the form

$$l \sim \frac{k^2}{B(0)} \tag{6}$$

where $B(0) = k^4(b-\varepsilon)^2 na^2 / \varepsilon^2$. Substituting this expression into Eq.(6) and taking into account that $k = \omega\sqrt{\varepsilon}/c$, we have

$$l \sim \frac{\varepsilon}{\frac{\omega^2}{c^2}(b-\varepsilon)^2 na^2}. \tag{7}$$

Substituting Eq.(7) into Eq.(3) one can be confirmed that $I_D \sim \varepsilon^{-3}(\omega)$. Therefore the radiation intensity enhances in the wavelength region where $\varepsilon(\omega) \ll 1$. Remind that $\varepsilon$ is the average dielectric constant of the system which for a layered stack has the form:

$$\varepsilon(\omega) = nab(\omega) + (1-na)\varepsilon_0(\omega).$$

Here $\varepsilon_0$ is the dielectric constant of a homogeneous medium into which plates with dielectric constant $b(\omega)$ and thickness $a$ are randomly embedded. If a homogeneous medium is vacuum then $\varepsilon_0 \equiv 1$. Choosing for plates materials with $b(\omega) < 0$ one can make the average dielectric constant of the system quite small $\varepsilon \ll 1$. Correspondingly, the photon elastic mean free path will be small and the radiation intensity will be large in a such system.

### 3. Results and Discussions

Let us now consider some specific examples. For dielectric constant of a simple metal one can use plasma formula $b(\omega) = 1 - \omega_p^2/\omega^2$, where $\omega_p$ is the plasma frequency. Therefore for frequencies $\omega < \omega_p$ the dielectric constant will be negative. Plasma frequency for simple metals is of order 20 -100 eV therefore the region where dielectric constant is negative lies from extreme ultraviolet to far infrared. For realization of diffusive mechanism of radiation absorption should be weak. This means that the plates should be very thin less than the depth of skin layer of metals in order to the photons can penetrate through them. In the optical region the skin layer of metals is of order of several hundred angstroms.
Therefore making a stack with such thin plates and vacuum within them will be very dfficult. In the other hand such a situation can be realized when a charged particle slides over a rough metallic surface [23-27]. In this case the randomly located hills and valleys will serve as plates and vacuum spacings between them, respectively. The energy of charged particle should be enough for penetration of the system with inessential lost of its energy. A few MeV electron energies are enough for penetration mm thickness of material. Let us estimate the number of emitted infrared photons for a stack of 50 plates with average thickness of $20 \mu m$ and average distance between plates $200 \mu m$.

In alkali halide crystals, in semiconductors like GaP; InSb; and etc. the dielectric constant is negative in the region between the frequencies of transversal and longitudinal optical phonons, see [28]. For example, for the compound MgO in the frequency region 550 - 650cm$^{-1}$ the real part of dielectric constant takes values in the interval  - 6; - 2 and the imaginary part in the interval  0.6 – 0.2. The above mentioned interval lies in the far infrared region. It follows from Eq.(7) that in case $2\pi a \leq \lambda$, the minimum of mean free path and therefore the maximum of radiation intensity is achieved for average plate thickness $a \sim \lambda/2\pi$. For the above mentioned frequencies this is about 20 $\mu$ m. Choosing such values for average thickness of plates one can reach the localization limit $\lambda/2\pi$ [29,30] for photon mean free path $l$. Note that the Eq.(3) is correct in the weak scattering diffusive regime $l \gg \lambda/2\pi$. Remind that an electromagnetic wave is localized provided that $l \leq \lambda/2\pi$. The above mentioned value for plate thickness 20 $\mu$ m is feasible and one can make a stack of such plates. Such a system could serve as s good source for far infrared radiation. One needs $l_{in}$ in order to estimate the number of emitted using Eq.(3). The inelastic mean free path of the photon in a random stack can be estimated as follows:

$$l_{in} \sim \frac{\lambda\sqrt{\varepsilon}}{\pi f \, \mathrm{Im} b(\omega)} \qquad (8)$$

where $f$ is the fraction of plates in the system. Taking $f \sim 0.1$, $\mathrm{Im} b \sim 0.4$ and $\varepsilon \sim 0.5$, one gets $l_{in} \sim 557 \mu m$. Using Eq.(3) one can estimate the integrated over all angles number of emitted photons in the interval $\Delta\omega$ as

$$N_{ph} \sim \frac{20}{3}\alpha \left(\frac{l_{in}}{l}\right)^2 \frac{\Delta\omega}{\omega} \qquad (9)$$

where $\alpha$ is the fine structure constant. Because $l \ll l_{in}$ the exponential decaying factor in Eq.(3) plays important role only for very large angles $\theta \approx \pi/2$. Therefore we ignored it when estimating the total number of emitted photons. Substituting $l_{in} \sim 562 \mu m$, $l \sim \lambda/2\pi \sim 17 \mu m$ into Eq.(8) and taking $\Delta\omega \sim \omega$ one has approximately $N_{ph} \sim 167$ infrared photons per one electron. This implies that using commercially available 5 - 6MeV , 1mA linear accelerator a total output power $2.4 mW (10^{18} \, photon/s)$ can be produced.

## 4. Conclusions

The generation of diffusive radiation by a charged particle passing through a random stack of plates in the infrared region is considered. We have shown, that for radiation intensity enhancement one needs to make the multiple scattering of pseudophotons on the plates more effective. For this goal we have suggested to use materials with negative dielectric constant. The output power is estimated. It was shown, that using commercially available $5-6\,MeV$, $1\,mA$ linear accelerator a total output power $2.4\,mW$ ($10^{18}\,photon/s$) can be produced.